\documentclass[aip,apl,preprint,endfloats,groupedaddress,superscriptaddress]{revtex4-1}
\usepackage{graphicx}

\begin{document}

\title{
Medium Energy Ion Scattering of Gr on SiC(0001) and Si(100)
}
\author{M.\ Copel }
\email[Electronic mail:] {mcopel@us.ibm.com}
\author{S.\ Oida}
\affiliation{IBM Research Division, T. J. Watson Research Center\\ P.O. Box 218, Yorktown Heights, NY 10598}
\author{A. Kasry}
\affiliation{IBM Research Division, T. J. Watson Research Center\\ P.O. Box 218, Yorktown Heights, NY 10598}
\affiliation{ Egypt Nanotechnology Center (EGNC), Smart Village, Giza, 12577 Egypt.}
\author{A. A. Bol}
\author{J. B. Hannon }
\author{R. M. Tromp}
\affiliation{IBM Research Division, T. J. Watson Research Center\\ P.O. Box 218, Yorktown Heights, NY 10598}

\date{\today}

\begin{abstract}
Depth profiling of graphene with high-resolution ion beam analysis is
a practical method for analysis of monolayer thicknesses of graphene. Not only
is the energy resolution sufficient to resolve graphene from underlying SiC, but by
use of isotope labeling it is possible to tag graphene generated from reacted ethylene. 
Furthermore, we are able to analyze graphene supported by oxidized Si(100) substrates,
allowing the study of graphene films grown by chemical vapor deposition on metal and transferred to silicon. 
This introduces a powerful method to explore the fundamentals of graphene formation.

\end{abstract}

\maketitle

Characterization of graphene formation on surfaces is an experimental challenge that
is pivotal to developing new electronic devices. 
Due to the extreme thinness of two-dimensional carbon layers, conventional
tools of solid-state physics are hard pressed to address structural issues
that are presented. Lateral coherence can be studied with probes 
such as transmission electron microscopy \cite{hashimoto_direct_2004}, scanning tunneling microscopy \cite{ishigami_atomic_2007} or low energy electron microscopy\cite{TrompPRL}, 
and structural information can also be garnered with Raman scattering\cite{Raman}.
However, these techniques are not generally used to learn about buried interfaces.
Medium energy ion scattering (MEIS) offers a unique capability of depth profiling
with isotopic specificity that can be used to study growth modes in a quantitative manner.
In this letter we will demonstrate that medium energy ion scattering (MEIS)
can be applied to graphene, opening up an avenue for research into challenging questions
concerning this new electronic material. MEIS has already been successfully applied to
studying the diffusion of oxygen through graphene to electronically decouple graphene
from the SiC substrate\cite{oida_decoupling_2010}.

MEIS has has been widely applied as a probe of thin films and surfaces, offering quantitative
analysis based on well understood Rutherford scattering\cite{CopelReview,gustafsson_high-resolution_2001}.
Results can be accurately
modeled with simulations based on independently determined parameters such as scattering
cross section and rate of energy loss in a host material\cite{grande_analytical_2007}.
However, to apply an ion beam probe to a graphene layer, several requirements must be met.
First, the sample must be macroscopic 
, since the probe samples an area of several square mm. 
Second, the background from the substrate must be suppressed enough for the carbon signal to be detected.
Both of these requirements can be met when the sample is Gr/SiC; large area samples
have been demonstrated on SiC substrates, and the use of a channeling geometry reduces the
substrate signal by two orders of magnitude. This means that ion beam analysis can at least
be attempted for Gr/SiC. Finally, we must consider whether there is sufficient resolution to discriminate
between surface Gr and the carbon in the SiC. The energy loss of a 100 keV proton traveling through carbon
is 13 eV/10$^{15}$ atoms/cm$^2$. The areal density of a monolayer (ML) of Gr is 3.9$\times 10^{15}$/cm$^2$. 
Including a geometric factor of 3.9 based on the angle of incidence (0 degrees) and exit angle (20 degrees),
the energy loss for each layer of graphene is 200 eV. Since MEIS has been demostrated
with an energy resolution of better than 150eV for 100keV protons, it is certainly feasible to
separate the backscatter peak of a multilayer graphene film from a SiC substrate.

\begin{figure}
\includegraphics[width=6.4in]{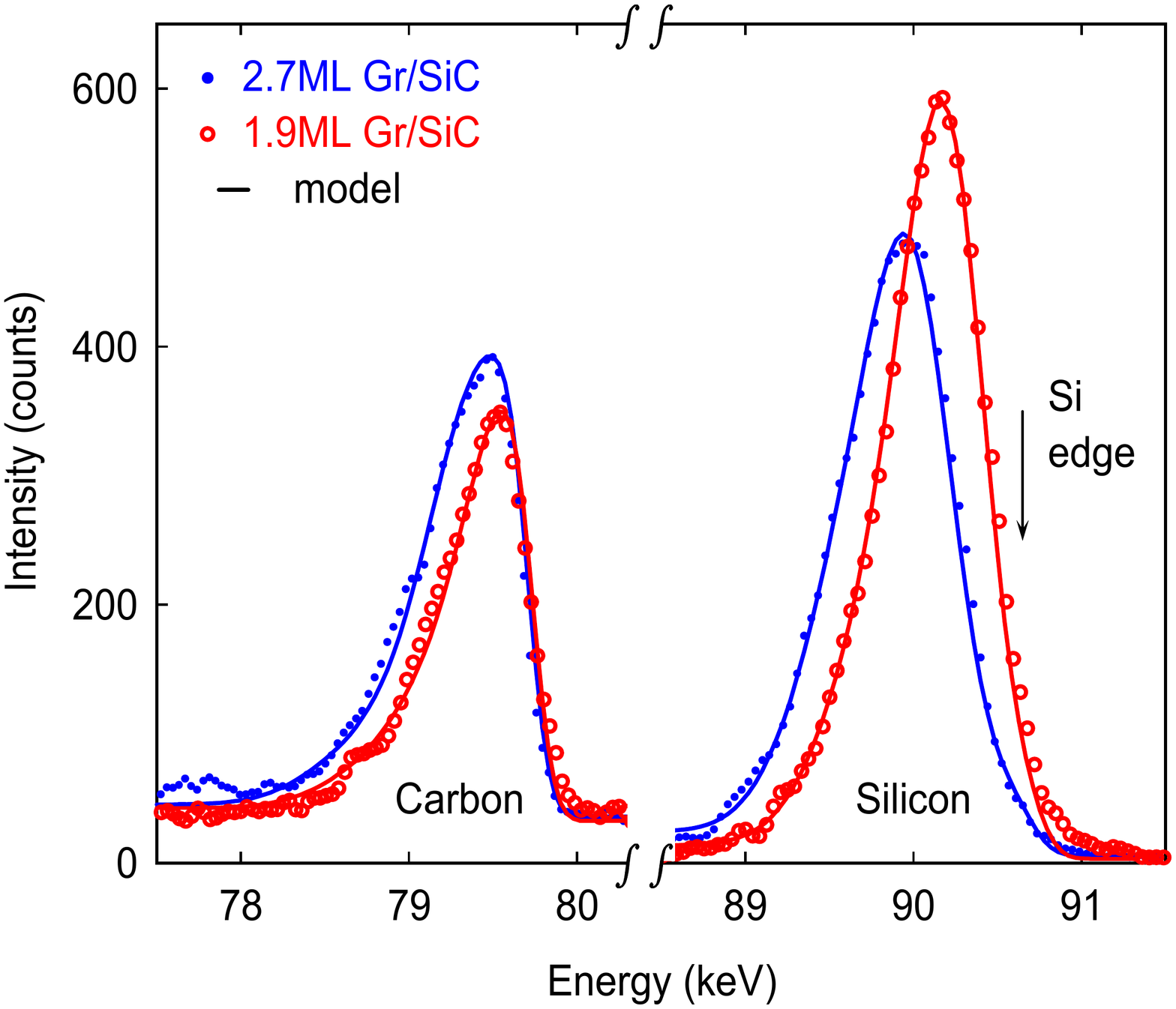}
\caption{
\label{sic03d2}
Backscattering spectra for two different thicknesses of Gr/SiC(0001). The carbon peak shows increased width for
the thicker film, demonstrating that the energy resolution is adequate to yield depth information. The silicon
peak is displaced toward lower energy for the thicker film, due to the stopping power of the graphene.
}
\end{figure}

As a test of applicability of MEIS to Gr/SiC, we examined two films with differing thicknesses
of graphene, both grown by sublimation of Si from Si-face SiC(0001) at high temperatures\cite{van_bommel_leed_1975}
in accordance with established techniques for generating smooth layers\cite{TrompPRL}. 
One sample was grown for 10 min. at 1180$^\circ$,
and a second sample was grown at about 20$^\circ$C higher temperature to generate a thicker Gr layer. 
After processing, the samples were transfered through air to the MEIS system.
Fig.~\ref{sic03d2} shows the carbon and silicon backscatter peaks from the two samples.
The film grown at higher temperature has a broader carbon peak with slightly higher intensity
compared to the low temperature film. The increased width of the peak is due to greater energy losses
experienced by protons as they traverse the graphene, indicating a thicker film.
An additional distinction can be seen in the position of the leading edge of the Si backscatter peak,
which occurs at lower energy for the thicker film. This is also an indication of a thicker graphene layer,
since the protons lose more energy going through the graphene to reach the underlying SiC, 
as well as reemerging to the vacuum.
Note that the top layers of the SiC are visible to the ion beam, since the graphene does not shadow the
underlying SiC. However, below the SiC surface peak the ions channel in crystalline SiC, reducing the yield
to a small background intensity. The Si peak is entirely due to the SiC, while the carbon peak contains contributions
from both Gr and SiC.
A quantitative model of the spectra, drawn as smooth curves in Fig.~\ref{sic03d2}, show that the samples
consist of 1.9 ML and 2.7 ML of graphene. In our models, the SiC contributes equal amounts to the Si and carbon peaks
after normalizing to the Rutherford cross section.
This result demonstrates that MEIS can not just discriminate between graphene layers of different thicknesses,
but has the capability of depth profiling, since the stopping power of an added layer of graphene
results in measurable shifts in spectral features.

\begin{figure}
\includegraphics[width=6.4in]{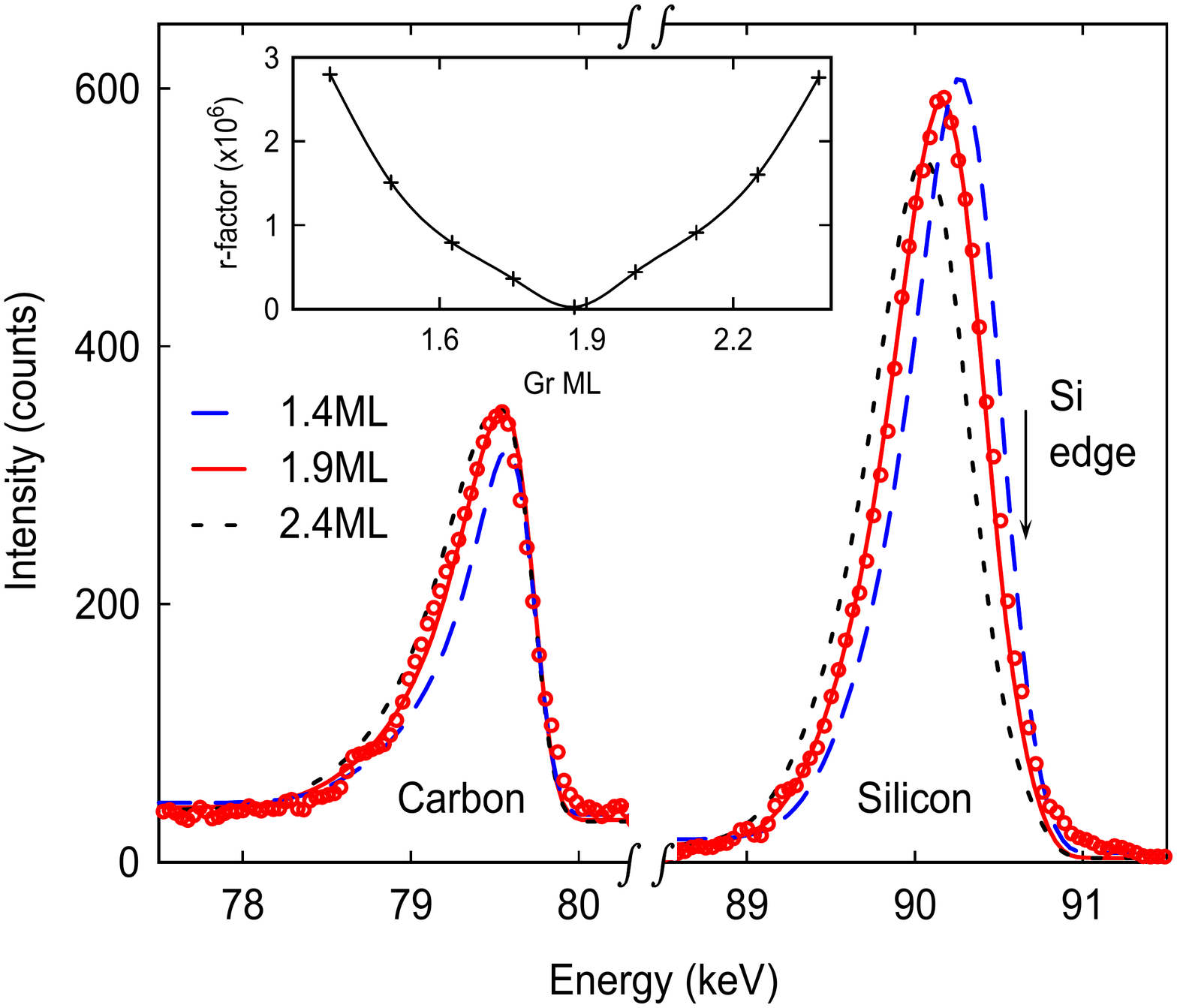}
\caption{
\label{sic03e}
Medium energy ion scattering spectrum of 1.9 monolayer Gr/SiC(0001), plotted with models for various thicknesses
of graphene. Inset shows goodness of fit, determined by r-factor analysis, as a function of graphene thickness.
}
\end{figure}

Simulations of MEIS results support the assertion that we can determine graphene thickness with submonolayer sensitivity. 
To illustrate this, we have plotted
a spectrum with 1.9 ML of graphene along with simulated results with coverages varying by 0.5ML (Fig.~\ref{sic03e}).
When the graphene thickness is altered from the optimum fit, large discrepancies occur for both
the carbon peak width and the location of the Si peak. The inset for Fig.~\ref{sic03e} shows the goodness of fit
based on an r-factor as a function of graphene thickness. A sharp minimum can be found near 1.9 ML, indicating
that deviations between experiment and simulation can be quantitatively assessed and used to guide
evaluation of data.

\begin{figure}
\includegraphics[width=6.4in]{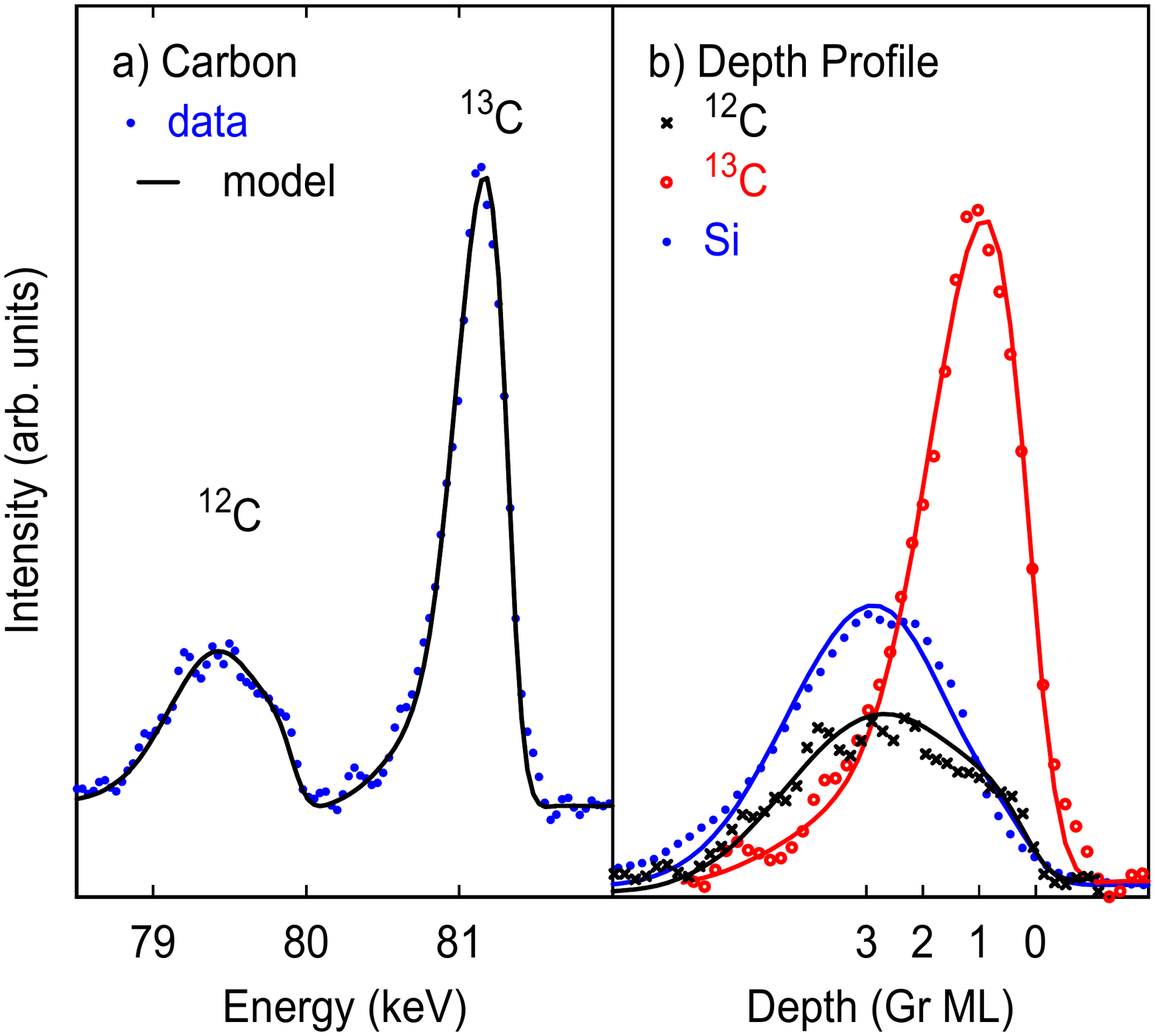}
\caption{
\label{sic15b}
Results for a SiC sample reacted with $^{13}$C$_2$H$_4$. (a) The carbon portion of the spectrum shows a
narrow peak due to surface $^{13}$C and a broad peak due to subsurface $^{12}$C. (b) Plotted on a depth scale,
the $^{13}$C signal is peaked at the surface, while the $^{12}$C and Si signals have their highest intensity
deeper in the sample.
}
\end{figure}

Additional information about graphene layers can be gathered by using $^{13}$C$_2$H$_4$ as a gas source.
Since scattering kinematics are dependent on target mass, protons backscattered from $^{13}$C have a
greater kinetic energy than protons scattered from $^{12}$C. This gives the capability of separating
a graphene layer generated by decomposition of $^{13}$C$_2$H$_4$ from the substrate signal, which should
be predominantly $^{12}$C.
In Fig.~\ref{sic15b}, we show a spectrum from a SiC sample that was cleaned by heating to 850$^\circ$C
in $1\times10^{-6}$ Torr disilane,
followed by exposure to $2\times10^{-5}$ Torr of $^{13}$C$_2$H$_4$ at 1140$^\circ$C.
After transfer through air, the sample was degassed for 1 min. at 750$^\circ$C to remove contaminants.
Two carbon peaks are visible in Fig.~\ref{sic15b}a: a sharp peak at 81.1 keV due to backscattering from surface $^{13}$C,
and a broad peak centered at 79.5 keV due to $^{12}$C in the top layers of the SiC.
We have plotted both carbon peaks on the same depth scale in Fig.~\ref{sic15b}b.
The $^{12}$C intensity peaks below the surface of the sample, since the SiC is mostly covered by graphene.
The smooth curves plotted on the data were generated from a model with 1.5 ML of graphene
consisting of 88\% $^{13}$C, covering 95\% of the surface. The model includes
32\% $^{13}$C (rel. to total C) in the top layers of the SiC.
We have succeeded in growing a graphene layer composed almost entirely of $^{13}$C covering
nearly the entire sample. It is not clear whether there is some small intermixing
of $^{13}$C with the underlying SiC, or whether our model underestimates the tail of the 
$^{13}$C signal. Nonetheless, the results establish that the use of carbon isotopes
combined with MEIS is a potential means for tracking the dynamics of graphene growth on SiC substrates.

Although this letter is mostly concerned with Gr/SiC, MEIS analysis is not limited to SiC. 
In fact, all that is required is a large area graphene film supported by a single-crystal substrate.
To illustrate this point, we turn to graphene films created by chemical vapor deposition (CVD)
on metal, followed by transfer to oxidized Si substrates\cite{li_large-area_2009}.
Following procedures similar to ref. \cite{kasry_chemical_2010},
a polycrystalline Cu foil was exposed to an ambient of 0.5 Torr of ethylene for 10 mins. 
Deposition was preceded by an {\it in situ} 10 min forming gas anneal at the growth temperature.
Afterwards, the sample was bonded to PMMA and the Cu was dissolved in FeCl$_3$.
The PMMA was removed after bonding to a Si sample with a native oxide present.
A proton beam is able to penetrate through the native oxide layer to channel in the crystalline Si,
reducing the background to the levels needed for carbon analysis by MEIS.
Consequently, we are able to gather data on Gr coverage as well as explore organic and inorganic
contamination. (The thick oxides typically used for optical selection of Gr flakes contribute too
much background intensity, and are unsuitable for MEIS.)

\begin{figure}
\includegraphics[width=6.4in]{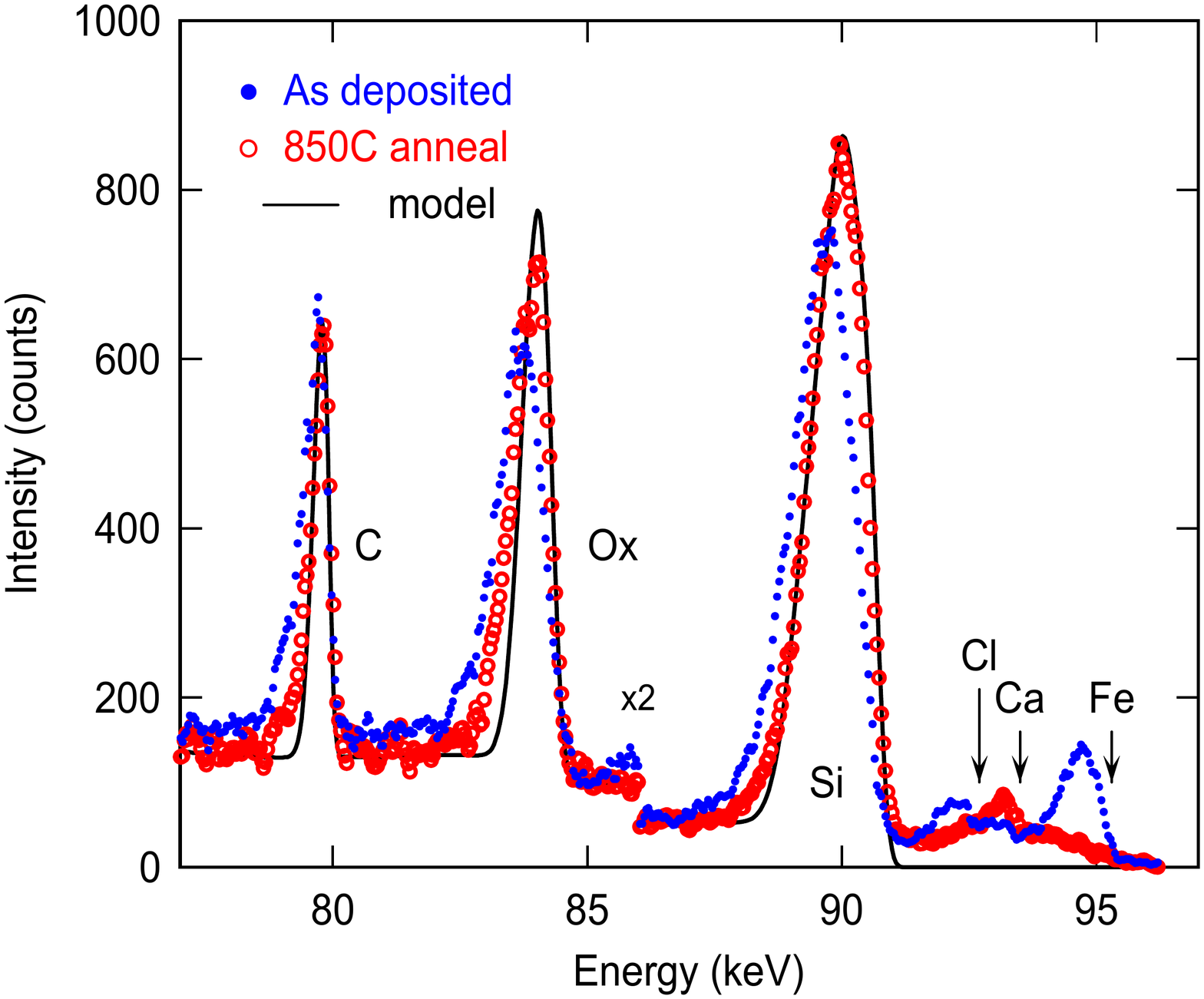}
\caption{
\label{sic42d}
Backscatter spectra for graphene films grown by chemical vapor deposition on Cu and transfered to Si.
Annealing removes carbon residue left over from the transfer process, sharpening the features.
Smooth curve is a model of an idealized graphene layer on an oxidized Si substrate.
}
\end{figure}

The freshly bonded wafer shows a carbon overlayer, in addition to silicon and oxygen peaks (Fig.~\ref{sic42d}).
At higher energy, Fe and Cl remnants can be observed.
Heating the sample to 850$^\circ$C sharpens the oxygen and silicon peaks, as well as reducing the Fe and Cl signals.
A peak from Ca is now apparent, possibly arising from impurities segregated from the Cu during 
the high temperature required for CVD.
Annealing not only improves the shape of the peaks, but it also removes some of the carbon and oxygen.
The volatile component is most likely organic residue left over from incomplete removal of the PMMA\cite{ishigami_atomic_2007}.
Prior to annealing, the sample contained $8.4\times10^{15}$/cm$^2$ carbon, 
and $7.7\times10^{15}$/cm$^2$ oxygen, which was reduced to
$6.5\times10^{15}$/cm$^2$ carbon and $7.4\times10^{15}$/cm$^2$ oxygen by annealing.
(1 ML of graphene is 3.9$\times10^{15}$/cm$^2$.)
Along with the data, we have plotted a simulated spectrum for Gr covering 90\% of the surface with a thickness of 1.4ML,
supported by 14\AA\ of SiO$_2$. Clearly, the experimental spectra are not as sharp as the model,
perhaps due to incomplete removal of contaminants or inhomogeneities in the samples.

It is important to note that the peak shapes and quantity of residual material are highly dependent on
sample preparation procedures. We are optimistic that improved film quality can be achieved by
honing processing conditions. The results serve to show that MEIS can be used to characterize
graphene transfered to Si(100) substrates, and provides a useful means of optimizing growth,
with rapid turn-around and minimal sample preparation.

The advances in synthesis of large area graphene films has made high-resolution ion beam analysis
a practical endeavor. This opens up the prospect of depth profiling experiments to learn about
growth modes by isotope tracing with atomic scale depth resolution. 
Experiments are not limited to SiC substrates, but are also feasible for films transfered to
single crystal substrates, such as silicon wafers.
We have used these capabilities to examine contamination issues that arise with organic polymers as support
during solvation of polycrystalline metal growth media.
Further experiments using MEIS will contribute to our ability to reproducibly fabricate
graphene electronic devices.


\bibliography{gr_meis}
\eject





\end{document}